\documentclass[traditabstract]{aa} 
\usepackage{graphicx}
\usepackage{txfonts}
\usepackage{natbib} 
\usepackage{mydef} 
\bibpunct{(}{)}{;}{a}{}{,} 
\begin{document}
   \title{GREAT [\ion{C}{II}] and CO observations of the BD+40$^\circ$4124 region}


   \author{G. Sandell \inst{1}
    \and   H. Wiesemeyer \inst{2} 
    \and  M. A. Requena-Torres \inst{2}
    \and  S. Heyminck \inst{2}
    \and  R. G\"usten \inst{2}
    \and  J. Stutzki \inst{3}
    \and R. Simon \inst{3}
    \and  U.  U. Graf \inst{3}
         }

   \institute{SOFIA-USRA, NASA Ames Research Center, MS 232-11, Building N232, Rm. 146, P. O. Box 1,
Moffett Field, CA 94035-0001, U. S. A.\\
              \email{Goran.H.Sandell@nasa.gov}
         \and
             Max Planck Institut f\"ur Radioastronomie, Auf dem H\"ugel 69, 53121 Bonn, Germany\\
           \and I. Physikalisches Institut der Universit\"at zu K\"oln, Z\"ulpicher Stra{\ss}e 77,  50937 K\"oln, Germany
             }

   \date{accepted February 7, 2012}

 
 \abstract{
  The BD+40$^\circ$4124 region was observed with high angular and
 spectral resolution with the German heterodyne instrument GREAT in CO J
 = $13 \to 12$ and [\ion{C}{II}] on SOFIA. These observations show that the
  [\ion{C}{II}]  emission is very strong in the reflection nebula
 surrounding the young Herbig Ae/Be star BD+40$^\circ$4124. A strip map over
 the nebula shows that the  [\ion{C}{II}]  emission approximately coincides with
 the optical nebulosity. The strongest  [\ion{C}{II}]  emission is centered on the
 B2 star and a deep spectrum shows that it has faint wings, which suggests
 that the ionized gas is expanding. We also see faint CO J = $13 \to 12$
 at the position of BD+40$^\circ$4124, which suggests that the star may
 still be surrounded  by an accretion disk. We also detected  [\ion{C}{II}] 
 emission and strong CO J = $13 \to 12$ toward V\,1318~Cyg. Here the  [\ion{C}{II}] 
 emission is fainter than in  BD+40$^\circ$4124 and appears to come from the outflow,
 since it shows red and blue wings with very little emission at the
 systemic velocity, where the CO emission is quite strong. It therefore
 appears that in the broad ISO beam the  [\ion{C}{II}]  emission was dominated by
 the reflection nebula surrounding  BD+40$^\circ$4124, while the high J
 CO lines originated from the adjacent younger and more deeply embedded binary system
  V\,1318~Cyg.}  
   \keywords{ISM: molecules --
              Stars: circumstellar matter  --
              Stars: pre-main sequence  --
              Stars:  variables: T Tauri, Herbig Ae/Be --
              (ISM:) photon-dominated region (PDR)
                               }

   \maketitle
%

\section{Introduction}
\begin{figure}[t] 
\begin{center}
\resizebox{\hsize}{!}{\includegraphics[angle=0,width=8.0cm,angle=0]{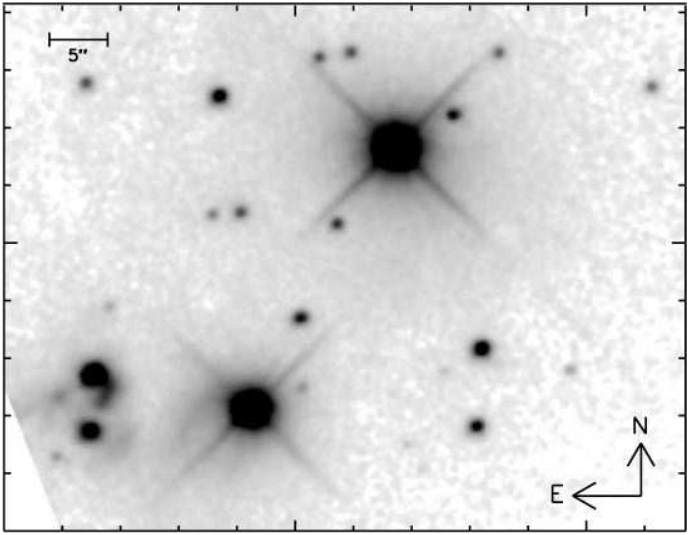}} 
\caption{A K-band image of the BD+40$^\circ$4124 field from
\citet{Davies01}.  BD+40$^\circ$4124 is the bright star in the northern
part of the image, while the second bright star {\bf 26\arcsec{}} to the south
east is V\,1686 Cyg. V\,1318~Cyg is the double star system at the south
eastern edge of the image. The southern star is the active one, showing
strong H$_2$ emission and powers a massive bipolar molecular outflow. \label{fig-kband} 
}
\end{center} 
\end{figure}

 The BD+40$^\circ$4124, at a distance of 980 pc \citep{Shevchenko91}, forms a 
 a small pre-main-sequence 
 cluster \citep{Herbig60, Hillenbrand95}. \citet{Herbig60}
 included BD+40$^\circ$4124 in his original list of Herbig Ae/Be (HAEBE) stars
 and noted that it had  three companions: LkH$\alpha$ 224 (V\,1686 Cyg),
 LkH$\alpha$ 225 (V\,1318 Cyg), and LkH$\alpha$ 226 within 35\arcsec\ of
 BD+40$^\circ$4124.   Both BD+40$^\circ$4124 and V\,1686 Cyg are  HAEBE
 stars with spectral types of B2 Ve and A7 Ve, respectively
 \citep{Ancker98,Ancker00}. \citet{Hillenbrand95}, however, classified
 V\,1686 Cyg as B5 Ve. V\,1318~Cyg is a heavily obscured binary
 system with a separation of $\sim$ 5\arcsec\ \citep{Aspin94}. The
 southern member of this binary, V\,1318~Cyg\,S, is more deeply embedded
  with a visual extinction of 10 mag or more. Several studies indicate that 
  V\,1318 Cyg may have a hidden companion about 1\arcsec\ to the NE that
  powers the outflow and the H$_2$O maser \citep{Davies01,Marvel05}.
   The spectral type of
 V\,1318~Cyg is very uncertain, but based on analysis of the spectral
 energy distribution, both \citet{Aspin94} and \citet{Ancker00} argued
 that at least one of the stars must be an intermediate mass star.
 \citet{Hillenbrand95} gave spectral classifications of  mid-A to Fe for
 both components. A K-band image of the BD+40$^\circ$4124 field is
 shown in Fig.~\ref{fig-kband}.

 BD+40$^\circ$4124 illuminates a bright reflection nebula with a radius
of $\sim$ 30\arcsec\ \citep{Herbig60,Loren77,Hillenbrand95}, while 
V\,1318~Cyg excites an H$_2$O maser and drives a bipolar molecular
outflow \citep{Palla95}.   The
BD+40$^\circ$4124 region was studied in the infrared by several groups using both the
ISO SWS and LWS spectrometers
\citep{Ancker00,Creech-Eakman02,Lorenzetti02}. ISO  LWS spectra were
obtained toward BD+40$^\circ$4124, V\,1686 Cyg, and V\,1318 Cyg, even
though they are all covered with one pointing, since the beam width
for LWS is $\sim$ 80\arcsec. \citet{Creech-Eakman02} and
\citet{Lorenzetti02} attributed all the [\ion{C}{II}] and high transition CO (J
= $14 \to 13$ to J = $16 \to 15$) emission to BD+40$^\circ$4124, while
\citet{Ancker00} assigned about the same amount of [\ion{C}{II}] emission to 
BD+40$^\circ$4124 and V\,1318 Cyg, and  no emission at all to V\,1686
Cyg.
 
 
In this letter we  revisit the BD+40$^\circ$4124 region with the
GREAT\footnote{GREAT is a development by the MPI f\"ur Radioastronomie
and the KOSMA/ Universit\"at zu K\"oln, in cooperation with the MPI f\"ur
Sonnensystemforschung and the DLR Institut f\"ur
Planetenforschung.}heterodyne instrument on SOFIA to see where the  [\ion{C}{II}] 
 and high J CO emission really comes from. Is it dominated by the bright
reflection nebulosity illuminated by BD+40$^\circ$4124 or does the
emission come from the younger, embedded star V\,1318~Cyg?   Given 
the low spatial resolution of the ISO data, it is clear that higher resolution 
data are needed to determine where the  [\ion{C}{II}] and CO emission comes from and which physical mechanisms generate this emission.

\section{Observations}

BD+40$^\circ$4124 and V\,1318~Cyg were observed with GREAT on SOFIA on
April 6, 2011 on a 53 minute leg at an altitude of  43,000 ft.  GREAT is a modular heterodyne instrument, with two
channels that are both used simultaneously. For a more complete description of
the instrument, see \citet{Heyminck12}. For this flight, which was the
first  science flight with GREAT, the configuration was set to the low-frequency
channel 1b (L1\#b), which covers the frequency range 1.42 - 1.52 THz,
and the low-frequency channel 2 (L2), covering 1.82 - 1.92 THz.  The L1\#b
channel was tuned to CO J =  $13 \to 12$ (1496.923 GHz) in the lower
sideband and the L2 channel was centered on the [\ion{C}{II}] $^2{\rm P}_{3/2} \to\  ^2{\rm P}_{1/2}$ (1900.5369
GHz) in the upper sideband. The half power beam width (HPBW) is $\sim$ 21\arcsec\  at 1.5 THz
and 16\arcsec\ at 1.9 THz.  As backend we used the fast  Fourier
transform spectrometers (AFFTS), which have a bandwidth of 1.5 GHz and
8192 channels and provide a frequency resolution of 0.212 MHz. The
measured system temperatures were $\sim$ 4,500 K for  [\ion{C}{II}] and
3,000 K for CO =  $13 \to 12$. All observations were made in dual beam
switch mode using a 150\arcsec\ chop amplitude in equatorial coordinates
at a position angle of 135\degr. We took ``long'' integration spectra
toward both BD+40$^\circ$4124 and V1318 Cyg with total integration
times of 2.5 and 3.7 minutes, respectively. We also obtained a strip map over
BD+40 in RA with 15\arcsec\ spacing going from -15\arcsec,0\arcsec\ to
+45\arcsec,0\arcsec\  to explore how extended the  [\ion{C}{II}]
emission was. The integration time for these spectra was 0.6
minutes/position. The spectra  were analyzed in CLASS, where we coadded
individual spectra and  removed a linear baseline. All spectra are calibrated in T$_{\rm A}^*$ and corrected for a
forward scattering efficiency of 0.95.

\section{Results}

\begin{table}
\begin{center}
\caption{Gaussian fits to long integration spectra}
\label{tbl-1}
{\scriptsize
\begin{tabular}{llrrcc}
\hline
\hline
Source & Line & $\int T_{\rm A}^{*} dV$ & $T_{\rm A}^{*}$ & $\Delta V$ & $V_{\rm LSR}$\\
       &    &   [K km s$^{-1}$]& [K] & [km s$^{-1}$]& [km s$^{-1}$] \\
\hline
BD+40$^\circ$41247 &[\ion{C}{II}] &    21.7 $\pm$ 1.7 &  9.9 &  2.1 $\pm$ 0.1  &  7.3 $\pm$ 0.1 \\
                &     &   23.5  $\pm$ 1.8 &    3.8 &  5.9 $\pm$ 0.4  &  7.0 $\pm$ 0.1  \\
     &CO (13-12)          &    2.1   $\pm$  0.3 &    0.7 & 2.5   $\pm$ 0.7 &  7.8 $\pm$ 0.2   \\
\hline     
V1318 Cyg   &   [\ion{C}{II}] &  17.2 $\pm$ 0.5 &  ... &   ...  & \\  
                       &     CO (13-12)  &  4.0 $\pm$ 1.0  & 1.2          &  \phantom{3}3.3 $\pm$ 0.5     &   8.1 $\pm$ 0.1    \\
                       &                           &  6.3 $\pm$ 1.0  & 0.5         &  11.7 $\pm$ 2.4  & 7.8 $\pm$ 0.6 \\
\hline
\end{tabular}
}
\end{center}
\end{table}


\subsection{BD+40$^\circ$4124}

\begin{figure}[t]
\begin{center}
\resizebox{\hsize}{!}{\includegraphics[angle=0,width=8.0cm,angle=-90]{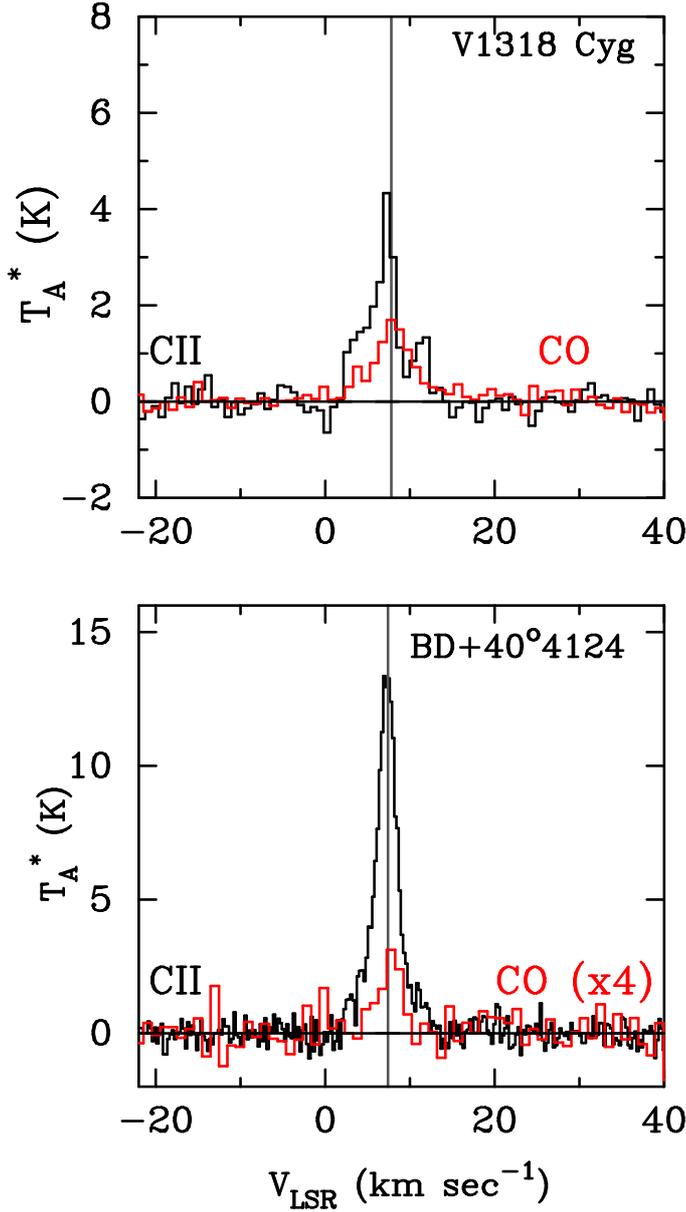}}
\caption{Deep  [\ion{C}{II}] and CO J = $13 \to 12$ spectra toward
BD+40$^\circ$4124 and V\,1318 Cyg.  [\ion{C}{II}] is plotted in black and the CO line profiles in
{\bf red}. For BD+40$^\circ$4124, where the CO J = $13 \to 12$ line is quite
faint, we multiplied the spectrum with a factor of four. The systemic velocity is marked by a vertical gray line. \label{fig-spectra}
 }  
\end{center} 
\end{figure}

\begin{figure}[t]
\begin{center}
\resizebox{\hsize}{!}{\includegraphics[angle=0,width=8.0cm,angle=-90]{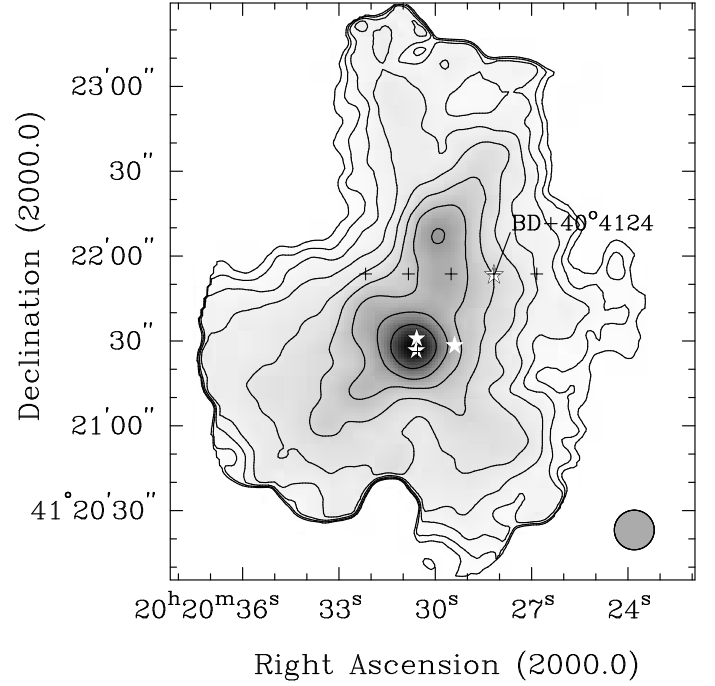}}
 \caption{850 $\mu$m continuum image of the BD+40$^\circ$4124 region from \citep{Sandell11}. The positions observed in
  [\ion{C}{II}] strip map are marked with black crosses.  BD+40$^\circ$4124 is indicated in the figure. 
  The two star symbols at the peak of the 850 $\mu$m contours are V\,1318 Cyg N and S, while the one to the west of them is V\,1686 Cyg.
 \label{fig-850}
 }
 \end{center} 
\end{figure}

\begin{figure}[ht]
\begin{center}
\resizebox{\hsize}{!}{\includegraphics[angle=0,width=8.0cm,angle=-90]{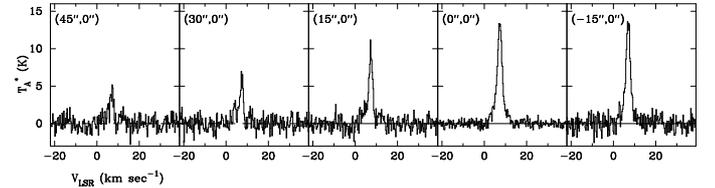}}
\caption{Strip map in RA over BD+40$^\circ$4124 in [\ion{C}{II}] 
with 15\arcsec\ spacing. The map suggests that [\ion{C}{II}] peaks slightly
west of the star, although the  offset could be caused by the pointing. 
 \label{fig-strip}
}
 \end{center} 
\end{figure}

The [\ion{C}{II}] emission is quite strong toward BD+40$^\circ$4124
with an antenna temperature of T$_{A}$ $\sim$ 13.5 K
(Fig.~\ref{fig-spectra}, Table~\ref{tbl-1}).  In Fig.~\ref{fig-850} we show the positions of the strip map 
that we obtained of  BD+40$^\circ$4124 marked as crosses  on the 850 $\mu$m SCUBA
image from \citet{Sandell11}. The strip map (Fig.~\ref{fig-strip})  shows that
the [\ion{C}{II}] emission peaks within errors on the star and 
slowly decreases to the east. There  does not appear to be any interaction between the [\ion{C}{II}]
emission and the dense molecular ridge (see Fig.~\ref{fig-850}), which extends from
V\,1318~Cyg to the north past BD+40$^\circ$4124
\citep{Sandell11,Looney06}. There are no changes in the  [\ion{C}{II}]  
brightness as it crosses over the ridge, nor is there any change in radial 
velocity or line width,  suggesting that BD+40$^\circ$4124 is in the
foreground, and not directly connected to the molecular ridge. The
extent of the  [\ion{C}{II}]  emission with a full width half maximum
(FWHM) of  $\sim$ 60\arcsec,  or about the same size as the reflection
nebulosity surrounding the star, confirms that the line emission
originates from the reflection nebulosity, and is almost certainly
photo-ionized by the UV-emission from the young B2 star. The long
integration spectrum (Fig.~\ref{fig-spectra}) shows a broad, $\sim$ 6
km~s$^{-1}$ wide pedestal (Table~\ref{tbl-1}), indicating that the
ionized gas is slowly expanding outward.

The CO J = $13 \to 12$ emission, which was observed simultaneously  with
[\ion{C}{II}], was also detected, but only at position (0\arcsec, 0\arcsec{}) toward the star itself
(Figure~\ref{fig-spectra}, Table~\ref{tbl-1}). If the line indeed had a similar
strength in the surrounding reflection nebula, it should have been
detected when we took an average of  the three spectra east of the star in the
strip-map. However, there is no hint of CO emission in these positions in the strip-map, confirming
that the CO emission comes from  BD+40$^\circ$4124 proper, which therefore must be
surrounded by hot, molecular gas, presumably in an accretion disk.

\subsection{V\,1318 Cyg}

We only obtained one long integration spectrum toward V\,1318 Cyg
(Fig.~\ref{fig-spectra}). CO J = $13 \to 12$ was easily detected with
T$_{\rm A}^{*}$ $\sim$ 1.7 K. The  emission is dominated by faint, broad emission
extending from $\sim$  $-$5   km~s$^{-1}$ on the blueshifted side to
$\sim$ +22  km~s$^{-1}$ on the redshifted side. A two-component Gaussian
fit gives a FWHM of  3.3~km~s$^{-1}$ for the narrow component and  11.7
km~s$^{-1}$ for the wide pedestal (Table~\ref{tbl-1}). The broad component is clearly associated with
the outflow powered by V\,1318  Cyg \citep{Palla95} and dominates the CO
emission, i.e., the integrated line emission is $\sim$1.5 times higher than
the emission from the narrow component.  The radial velocity of the narrow component,  8.1 km~s$^{-1}$,
agrees well with the systemic velocity of the star, 7.9 km~s$^{-1}$  determined from high spatial
resolution  CO J = $2 \to 1$ observations \citep{Looney06} and is likely to
originate in the accretion disk or the surrounding envelope.

In contrast, the  [\ion{C}{II}] emission looks quite different. At  first glance it looks as if the
spectrum is self-absorbed. However, there is no reason why the [\ion{C}{II}] would be affected by self-absorption.  
Instead it appears that the emission is completely dominated
by the outflow. Even though the CO J = $13 \to 12$ line profile suggests
that the blueshifted wing is somewhat stronger than the redshifted
one, the difference is quite marginal, while the blueshifted emission
completely dominates in [\ion{C}{II}]. Most of the emission is at low
 velocities and we see no ``high-velocity'' gas in [\ion{C}{II}] .
Since we have not mapped the emission, we cannot say how extended it is.
There is probably some [\ion{C}{II}]  emission throughout the outflow
lobes, although the [\ion{C}{II}]  is likely to be stronger close the
exciting star, see the discussion section below.

\section{Discussion}


\subsection{ [\ion{C}{II}] emission}

Our observations of  [\ion{C}{II}] in the BD+40$^\circ$4124 region show
that most of the  [\ion{C}{II}] emission comes from the reflection
nebulosity surrounding the B2 star. Based on our strip map,
we estimate the [\ion{C}{II}] emission to have a FWHM of $\sim$
60\arcsec\, similar to that of the reflection nebula.  The integrated
line intensity from  BD+40$^\circ$4124 is  therefore $\sim$ 40 $\times$
10$^{-15}$ W~m$^{-2}$ assuming a conversion factor of 1000 Jy/K  between
antenna temperature (T$_A^*$) and flux density. This estimate is very
uncertain  ($\sim$ 50\%) because of the uncertainty in the size of the
emitting region, but it does show that almost  the entire [\ion{C}{II}] 
emission observed by ISO comes from the reflection nebula illuminated by
BD+40$^\circ$4124. \citet{Lorenzetti02} and \citet{Creech-Eakman02}
reported a  [\ion{C}{II}] line intensity of 56  $\times$ 10$^{-15}$
W~m$^{-2}$ toward BD+40$^\circ$4124. This agrees well with what 
\citet{Ancker00} found for  BD+40$^\circ$4124, although they assigned a
similar line intensity to  V\,1318 Cyg. Even though we did not 
map the region around V\,1318 Cyg, it is quite clear from our [\ion{C}{II}]
spectrum of V\,1318 Cyg (Figure~\ref{fig-spectra}, Table~\ref{tbl-1})
that the emission is dominated by the outflow.  The line intensity toward V\,1318 Cyg is
only  about a third of what we see toward  BD+40$^\circ$4124. 

There are very few observations of outflows in [\ion{C}{II}] that have
both sufficient spatial and/or spectral resolution  to investigate the source of the  [\ion{C}{II}]  emission. The jet-like
low-mass outflow HH\,46 was observed with the integral field
spectrometer PACS centered on the low-mass protostar \citep{van
Kempen10}.  These observations show that the  [\ion{C}{II}] emission is
about twice as strong in the blueshifted outflow lobe compared to the
redshifted outflow lobe or the central protostar. Since there is no
velocity information in these data, it is not clear whether the
emission comes from low-velocity UV-heated gas in the cavity walls or
whether [\ion{C}{II}] is mixed with the molecular high-velocity gas.
Unfortunately, there were no observations far away from the central
source, which makes it difficult to check whether [\ion{C}{II}] requires
direct UV-excitation or whether C-shocks in the outflow would be equally
efficient. The similar line strength throughout the red- and blueshifted
outflow lobes suggests that in a low-mass outflow we mostly see emission
from the cavity walls, which would agree with what we see toward
V\,1318~Cyg with GREAT, where we only see [\ion{C}{II}] at low
velocities.  HIFI observations of the massive DR\,21 outflow, however,
show [\ion{C}{II}] emission that is as broad as the molecular line
emission \citep{Ossenkopf10}. DR\,21, however, is a high-mass star
formation region with intense UV-emission.

The size of the outflow powered by V\,1318 Cyg is $\sim$  55\arcsec\ \citep{Palla95}. If we
assume that the [\ion{C}{II}] emission is similar in strength as what is
seen toward V\,1318 Cyg, we would expect something like 15 $\times$
10$^{-15}$ W m$^{-2}$. This confirms that the
reflection nebula surrounding BD+40$^\circ$4124 dominates the 
[\ion{C}{II}] emission in the large ISO beam with a minor contribution  from the V\,1318 Cyg
outflow.

\subsection{CO J = $13 \to 12$ emission}

The detection of CO J = $13 \to 12$ emission towards  BD+40$^\circ$4124
is intriguing. \citet{Sandell11} did not detect any submillimeter continuum
emission towards the star with an upper limit of 35 mJy beam$^{-1}$ at
850 $\mu$m, suggesting that the star may already have dispersed its
disk. \citet{Skinner93}, however, reported a marginal detection of
free-free emission at 3.6 and 6 cm, which, if real, could mean that
there is still some remnant disk around the star, since free-free
emission from early B-stars is generally believed to originate from
photoionization of circumstellar disks \citep{Hollenbach00}. Because we
only see hot CO toward the star, it is unlikely that it would come from
the surrounding envelope and it is therefore more plausible that it
originates in a circumstellar disk that surrounds the star. Such a disk
would have moderate inclination, because the extinction toward the star
is low, A$_V$ $\sim$ 3 mag \citep{Ancker98}, which agrees with the
relatively narrow linewidth of the CO $13 \to 12$  line, 2.5 km~s$^{-1}$
(Table~\ref{tbl-1}). Looking for high J CO emission may therefore be
a good way to probe the gas in disks around early B-stars, since the
gas is expected to be much hotter than in low-mass stars, owing to the much higher
UV-field that  illuminates and heats the disk. It would therefore be very valuable to
search for hot CO in the few early B-stars that show strong evidence for
a circumstellar disk, such as HD200775, which illuminates the reflection nebula
NGC\,7023 \citep{Okamoto09}, or MWC\,349, which has a largely ionized disk, see e.g. \citet{Sandell11}.

In contrast, the CO J = $13 \to 12$ emission from the deeply embedded V\,1318~Cyg is
much stronger than for BD+40$^\circ$4124, and dominated by the hot outflow,
not by the disk. The CO line has a broad pedestal with a FWHM of $\ge$ 10 km s$^{-1}$,
which comes from the outflow and contributes to more than half of the line intensity. There is also
a narrow component, which could come from the accretion disk, although it could also originate in the
dense envelope that surrounds the star.
 
\section{Conclusions}

Observations with GREAT on SOFIA toward the BD+40$^\circ$4124 group show that
the [\ion{C}{II}] emission seen by ISO is dominated by emission from the reflection nebula illuminated by 
BD+40$^\circ$4124. However, we do also see  [\ion{C}{II}] emission from the outflow powered
by V\,1318~Cyg, which is a deeply embedded young HAEBE star $\sim$ 35\arcsec\ southeast
of BD+40$^\circ$4124. On the other hand, the high J CO emission observed by ISO is completely dominated by
hot gas in the outflow from V\,1318~Cyg. We also detected faint CO J = $13 \to 12$ emission toward
BD+40$^\circ$4124, which suggests that the stars is still surrounded by an accretion disk.

\begin{acknowledgements}
Based  on observations made with the NASA/DLR Stratospheric
Observatory for Infrared Astronomy. SOFIA Science Mission Operations are 
conducted jointly by the Universities Space Research Association, Inc., under
NASA contract NAS2-97001, and the Deutsches SOFIA Institut under DLR
contract 50 OK 0901. We also thank  Hans Zinnecker for a critical reading of the paper.
 \end{acknowledgements}

\end{document}